# Enhanced Conductance Fluctuation by Quantum Confinement Effect in Graphene Nanoribbons

*Guangyu Xu[1*], Carlos M. Torres Jr.[1], Emil B. Song[1], Jianshi Tang[1], Jingwei Bai[2], Xiangfeng Duan[3], Yuegang Zhang[4*], Kang L. Wang[1]*

[1]Department of Electrical Engineering, [2]Department of Material Science and Engineering, [3]Department of Chemistry and Biochemistry, University of California at Los Angeles, Los Angeles, California 90095, USA

[4]Molecular Foundry, Lawrence Berkeley National Laboratory, 1 Cyclotron Road, Berkeley, California 94720, USA

* Correspondence: guangyu@ee.ucla.edu (GX); yzhang5@lbl.gov (YZ).

**ABSTRACT**. Conductance fluctuation is usually unavoidable in graphene nanoribbons (GNR) due to the presence of disorder along its edges. By measuring the low-frequency noise in GNR devices, we find that the conductance fluctuation is strongly correlated with the density-of-states of GNR. In single-layer GNR, the gate-dependence of noise shows peaks whose positions quantitatively match the subband positions in the band structures of GNR. This correlation provides a robust mechanism to electrically probe the band structure of GNR, especially when the subband structures are smeared out in conductance measurement.

**Keywords:** graphene nanoribbon, quantum transport, low-frequency noise, electrical probing, band structure.

Considerable interest in mesoscopic low-dimensional electron systems has been motivated by a variety of unusual transport phenomena originating from the quantum confinement effect[1-3]. Inherent to quantum transport systems, enhanced conductance fluctuations due to disorder have brought intensive discussions on their correlation with the quantized density-of-states[4,5]. To date, research on conductance fluctuations is not only the subject of main scientific concerns in kinetics of quantum transport[6,7], but also of particular interest in probing the electronic band structure of low-dimensional nanostructures[8,9]. Low-frequency conductance fluctuation, or the flicker noise, has been broadly utilized to characterize the trap-induced charge switching processes[10,11]. The finding of strong correlation between the trap-induced conductance fluctuations (noise) and the density-of-states (DOS) in low-dimensional nanostructures has raised broad interest in the research of two-dimensional electron-gas (2DEG) and multiple cross-fields[9,12-15].

Graphene has recently attracted much attention for its ultrahigh intrinsic carrier mobility and thermal conductivity, both of which greatly benefit device applications[16,17]. Graphene nanoribbon (GNR), the graphene with a nanometer-size width, is a quasi-one-dimensional (1D) material with the presence of an energy gap[18], which is advantageous over the large graphene (no bandgap) for switching on/off the devices. However, the quantum transport of GNRs is still poorly understood: much debate remains on the origins of the energy gap in GNR[19-22], whereas the correlation of conductance fluctuations with the quantum confinement effect along the width of GNR has been amazingly unexplored. On the other hand, although previous works report the subband formation of GNR as indicated in the observation of quantized conductance plateaus[23,24], these plateaus can be significantly disturbed by multiple types of disorder similar to the case in quantum wire systems[25-27]. A more robust and direct probing mechanism to the band structure of GNR is thus critical and of fundamental interest to examine the intrinsic electrical properties of GNRs. In this context, a recent work by Lin et al. showed the room temperature noise measurement of GNRs, which shed light on the noise dependence on the change of energy gap in bi-layer GNRs[28].

In this Letter, we report the observation of strong correlation between the enhanced conductance fluctuations of GNR and its electronic band structure. By measuring the low-frequency noise of GNR devices at low temperature (77K), we find that the enhanced conductance fluctuations (noise) originate from the quantum confinement effect along the GNR widths: in single-layer GNR (SLR), the gate-dependence of noise shows peaks whose positions quantitatively agree with the quasi-1D transport theory; in bi-layer GNR (BLR), the noise peaks are also obvious while the band structure is unclear in conductance data. The correlation between noise and DOS provides a robust mechanism to directly probe the band structure of GNRs, especially when the subband feature is smeared out in conductance measurement. Our result can extend to a broad range of low-dimensional nanostructures and re-invigorate interest towards implementing conductance fluctuations (noise) for fundamental studies of quantum transport systems.

GNR samples were prepared by a nanowire-mask based fabrication method as reported before[29]. The GNRs were patterned into multi-probe structures by e-beam lithography. GNR devices were maintained in vacuum environment, and a vacuum bakeout process was applied to partially desorb contaminants before the measurements[21]. We apply a standard four-probe setup to reduce the noise contribution from the contacts[30]: an Agilent 4156C was used to apply dc current to the GNR within its linear regime; an Agilent 35670A was used to collect the noise spectra of the fluctuations in the potential difference (V) across the GNR sample. Under each gate bias ($V_g$), the dc conductance (G) and low-frequency noise spectra ($S_V$) were collected at the same time to avoid the hysteresis effect (see Supplement). We use the frequency (f)-averaged noise figure $A = \frac{1}{Z}\sum_{i=1}^{Z} f_i \times S_{Vi}/V^2$ to characterize the noise level, where the noise spectra follow with a $1/f^\alpha$ behavior ($\alpha$ ranging from 0.85 to 1.12)[31]. The dimension of GNR samples (the length (L) and width (W)) are measured by detailed atomic-force-microscopy (AFM) scanning[29].

Fig. 1a shows the typical temperature (T)-dependent G versus $V_g$ curves (shifted by the gate bias at Dirac point, $V_{Dirac}$) for a SLR sample, where the electron-hole asymmetry may be caused by the contact-doping effect[32]. Due to quasi-1D subband formation, the conductance plateaus appear at T=77K for both electron- ($V_g$-$V_{Dirac}$>0) and hole-conduction ($V_g$-$V_{Dirac}$<0) sides, and smear out at T=300K[23]. The negative differential conductance region (i.e. $\partial G/\partial |V_g|$<0 as circled) results from the enhanced inter-subband scattering near the transition of conductance plateaus[27].

To see the noise correlation with quasi-1D subband formation, we rescale the gate-dependence of the noise as $\sqrt{|V_g-V_{Dirac}|}$ in SLR and $|V_g-V_{Dirac}|$ in BLR for the rest of this paper[33, 34]. This representation helps show the noise behavior in the energy scale ($E_{SLR} \sim \sqrt{|V_g-V_{Dirac}|}$ and $E_{BLR} \sim |V_g-V_{Dirac}|$), where the difference for SLR and BLR comes from their different energy dispersions[34]. We confirmed that the noise spectra follow 1/f behavior (see top inset of Fig. 1b) at each applied gate bias, thus the definition of noise figure (A) is valid[9, 28].

Fig. 1b shows the gate-dependence of the noise figure (A) for electron-conduction side of SLR1 (the data for hole-conduction side are qualitatively the same). At T=77K, the noise curves (top) show noise peaks near the transition of conductance plateaus (bottom). The main noise peaks (as arrowed) are nearly equidistant in the scale of $\sqrt{V_g-V_{Dirac}}$; this is consistent with the fact that the separation of the quasi-1D subbands (see the bottom inset of Fig 1b) is nearly constant (the K-K' subband splitting effect is discussed in Fig. 2) [23, 24, 27]. The noise peaks are confirmed to be independent from the applied current, gate sweeping direction and frequency range, assuring that their origin comes from intrinsic trapping/de-trapping processes near the SLR.

We propose that the noise peaks originate from the enhanced trap-induced conductance fluctuations which follow the density-of-states (DOS) of SLR (see bottom inset of Fig. 1b). Starting from a qualitative explanation, the correlation between the conductance fluctuations and the band structure of SLR can be viewed as follows: In the presence of trap states, the trapping/detrapping processes cause the fluctuations of Fermi energy ($\Delta E_F$) which originate from the change of screened Coulomb potential of the traps. Consequently, the $\Delta E_F$ induces enhanced conductance fluctuations near the threshold of the nth subband ($E_n$), because the DOS ($\rho_n(E) \propto (E-E_n)^{-1/2}$) diverges near $E \sim E_n$[27]. Correspondingly, the conductance fluctuations are suppressed between the thresholds of subbands where the profile of DOS is smoother. Our physical picture is consistent with the general noise model as referred in III-V quantum structures, $S_V/V^2 = S_G/G^2 \propto (1/G^2) \times (\partial G/\partial E_F)^2$, where $S_G$ is the noise spectral density of the conductance fluctuations[12]. We note that the noise peaks disappear at 300K and at higher gate biases, which mainly result from the subband mixing caused by the temperature and disorder (short-range), respectively[26, 27].

To quantify our explanation, we further study the dimension-dependence of the noise. Quantitative analysis shows that the noise peak is indeed a measure of the band structure as described below.

As W increases, the separation of noise peaks decreases (as arrowed in Fig. 2a); this fact is consistent with the W-dependence of subband separations in SLR[23, 26, 27]. We apply the classical model of a conducting strip to estimate the gate capacitances as[23, 35]: $C_g \sim C_C \sim 4.6 \times 10^{11} cm^{-2}V^{-1}$ (W~42nm, $t_{ox}$~322nm) and $3.2 \times 10^{11} cm^{-2}V^{-1}$ (W~68nm, $t_{ox}$~305nm). Consequently, using the energy relation $E = \hbar v_F \sqrt{C_g \pi |V_g - V_{Dirac}|}$, the measured noise peak separation (as arrowed) can be converted to be $\Delta E$~82meV (W=42nm) and 49meV (W=68nm). Based on the band structure, the subband separation is $\Delta E_{DOS} \sim \pi \hbar v_F/W$ ~49meV (W=42nm) and 30meV (W=68nm), respectively. The discrepancy between $\Delta E$ and $\Delta E_{DOS}$ (within a factor of 2) is due to the overestimation of $C_g$[23, 35]. The ratio $\Delta E_{DOS}(42nm)/\Delta E_{DOS}(68nm)$ ~1.62 reasonably matches the ratio $\Delta E(42nm)/\Delta E(68nm)$ ~1.68 for the noise peak separations; this fact strengthens our statement that the noise peaks strongly correlate to the subband structures formed by the quantum confinement along the width.

As L increases, the number of noise peaks decreases (as arrowed in Fig. 2b) because of the subband mixing resulting from an increase of disorder.

In Fig. 2b, we also observe a series of seven noise peaks in SLR3 (W/L~30nm/0.61μm) which quantitatively follow the K-K' subbands splitting in the DOS. That is, the noise peaks show alternative separations, giving $\Delta E'/\Delta E \sim 0.29$ ($\Delta E'$ is the K-K' subband splitting), a ratio comparable to theoretical predictions for armchair nanoribbons (0.13-0.46) [23, 24]. The ratio $\Delta E(30nm)/\Delta E(42nm) \sim 1.9$ ($\Delta E(30nm)$ is estimated between the odd-order noise peaks) reasonably matches the ratio $\Delta E_{DOS}(30nm)/\Delta E_{DOS}(42nm) \sim 1.4$. This K-K' subband splitting effect is not clearly observed in SLR1 and SLR2 (Fig. 2a), where the splitting separation ($\Delta E'$) can be smeared out by temperature. To prove this, we did measurement for SLR1 at T=4.19K, where extra noise peaks (corresponding to the K-K' subband splitting) are observed ($\Delta E'/\Delta E \sim 0.4$) comparing with the case at 77K (see Supplement).

While the noise peaks are well defined by the quantum confinement along the width, we note that the ideal conductance plateaus can be significantly disturbed by disorders (see Fig. 2a and 2b). For example, SLR2 (W~68nm) shows lower conductance than that of SLR1 (W~42nm), indicating a smaller transmission coefficient of subbands in SLR2 due to disorder[23, 27]. Detailed AFM scanning shows that SLR2 is indeed more disordered (e.g. non-uniformity of width and chemical residual, see Supplement). The reason may be that while all disorders contribute to the scattering that disturb the conductance from the ideal case, only the trap states near the Fermi-energy (partial disorder) that cause the trapping-detrapping processes contribute to the measured noise. This fact suggests that the noise is a more direct probing mechanism for the subband structure of SLRs than the conductance.

Comparing with SLR, BLR has more complicated subband structures and the theoretical works on its fundamental transport are still rare[36, 37]. We thus limit the discussion of BLR to be qualitative.

The main feature for BLR is that (see Fig. 3): the noise peaks are apparent at T=77K (smeared at 300K), whereas the corresponding conductance plateaus are not as obvious. Also, BLR shows more noise peaks than SLR, and the noise peaks (as arrowed) are not equidistant in the scale of $V_g-V_{Dirac}$; both facts are consistent with the BLR subband structure, where the separation of subbands is smaller than SLR and not equidistant in the energy scale[36]. Thus, similar to the discussions in SLR, we attribute these noise peaks in BLR to the enhanced conductance fluctuation near its subband thresholds. This fact reaffirms that the noise can act as a robust mechanism to directly probe the DOS, especially for BLR whose subband structures are smeared out in conductance measurement. Complicated by the subband structure, we did not see clear dimension-dependence of the noise peaks in BLR[38]. Besides, we note that the noise level of BLR is overall lower than that of SLR with similar L and W, which may relate to the carrier screening effect in BLR[28].

Finally, we discuss about the T-dependence of the noise behavior in our GNRs. In our measurement, no clear (or weak) T-dependence of conductance value is observed for both SLR and BLR (see Supplement); while the noise floor (other than the noise peaks) is larger at 77K than 300K (see Fig. 1b and Fig. 3). This fact appears similar to the noise theorem in the hopping regime[39, 40]: the noise increase at 77K can come from the reduced number of conducting paths, thus the effective conducting area reduces and the noise (~1/Area) increases. Furthermore, we measured the large graphene (~μm width) and found an opposite T-dependence: the noise is lower at 77K than 300K (not shown). This suggests that the edge significantly impacts the noise behavior of GNRs, since edge states are more influential in GNR transport than large graphene.

In conclusion, we observe the enhanced conductance fluctuations in GNRs induced by the quantum confinement effect along the widths; this is clearly proved by the strong correlation between the noise peaks and the band structure (DOS) of the GNRs on the quantitative level. This correlation provides a robust mechanism to electrically probe the band structure of GNRs, especially when the subband feature is smeared out in the conductance (G) measurement. Our result can extend to research in a broad range of low-dimensional nanostructures, which promotes the noise-based metrology for the electronic band structure of quantum transport systems.

**ACKNOWLEDGMENT** The authors thank for helpful discussions with F. Miao, Y. Zhou, X. Zhang and I. Ovchinnikov and experimental support from S. Aloni and T. Kuykendall. This work was in part supported by MARCO Focus Center on Functional Engineered Nano Architectonics and the U.S. Department of Energy under Contract No. DE-AC02-05CH11231.

**FIGURE CAPTIONS**

Figure 1 **Temperature-dependence of noise behavior (and conductance) in single layer graphene nanoribbon (SLR). a.** T-dependent dc conductance (G) versus gate bias ($V_g$-$V_{Dirac}$) for SLR1 (W~42 nm, L~0.81μm). The circled area shows negative differential conductance regime at 77K. The inset shows the AFM image of the measured SLR1. The scale bar is equal to 0.5 μm. **b.** T-dependent noise (A, top) and conductance (G, bottom) in the electron-conduction side of SLR1 in scale of $\sqrt{V_g - V_{Dirac}}$. At T=77 K, the noise peaks (top) are shown to appear (not apparent at T=300 K) near the transition of the conductance plateaus (bottom) near the Dirac point (as arrowed), which suggest their correlation with the quasi-1D transport in SLR. Vertical dotted lines are guides to the eyes. The upper inset shows typical 1/f noise spectra collected in SLR1, where the background noise measured at zero current bias

has been subtracted. The solid line shows the normalized 1/f noise scaled to $S_V = 1 \times 10^{-8} V^2/Hz$ at $f \sim 1 Hz$. The bottom inset shows the schematics of quasi-1D DOS in scale of $\sqrt{V_g - V_{Dirac}}$ (The K-K' subband splitting, $\Delta E'$, is not in scale).

**Figure 2 Dimension-dependence of noise behavior (and conductance) in single layer graphene nanoribbon (SLR). a.** W-dependence of noise behavior of SLR1 (W~42 nm) and SLR2 (W~68 nm) with a similar length (L~0.81 μm) at T=77 K in scale of $\sqrt{V_g - V_{Dirac}}$. As W increases, the separation of main noise peaks (as arrowed) near the Dirac point decreases. The inset shows the effective gate capacitance of SLRs, $C_{total}$, versus the width (W) using the classical model of a conducting strip (infinitely-long). The oxide thicknesses, $t_{ox}$, are measured as: $t_{ox}$(W~30nm)~$t_{ox}$(W~42nm)~322nm, $t_{ox}$(W~68nm)~305nm. **b.** L-dependence of noise behavior of SLR3 (L~0.61 μm) and SLR4 (L~1.13 μm) with a similar width (W~30 nm) at T=77 K in scale of $\sqrt{V_g - V_{Dirac}}$. As L increases, the number of noise peaks (as arrowed) decreases, possibly because the disorder results in subband mixing. The inset shows the schematics of quasi-1D DOS in scale of $\sqrt{V_g - V_{Dirac}}$.

**Figure 3 Temperature-dependence of noise behavior (and conductance) in bilayer graphene nanoribbon (BLR).** T-dependence of noise behavior in BLR1 (W~49 nm, L~0.79 μm) is presented in scale of $V_g - V_{Dirac}$. At T=77 K, the noise peaks (as arrowed) appear while the corresponding conductance plateaus (bottom) are not obvious.

Figure 1
Xu et. al.

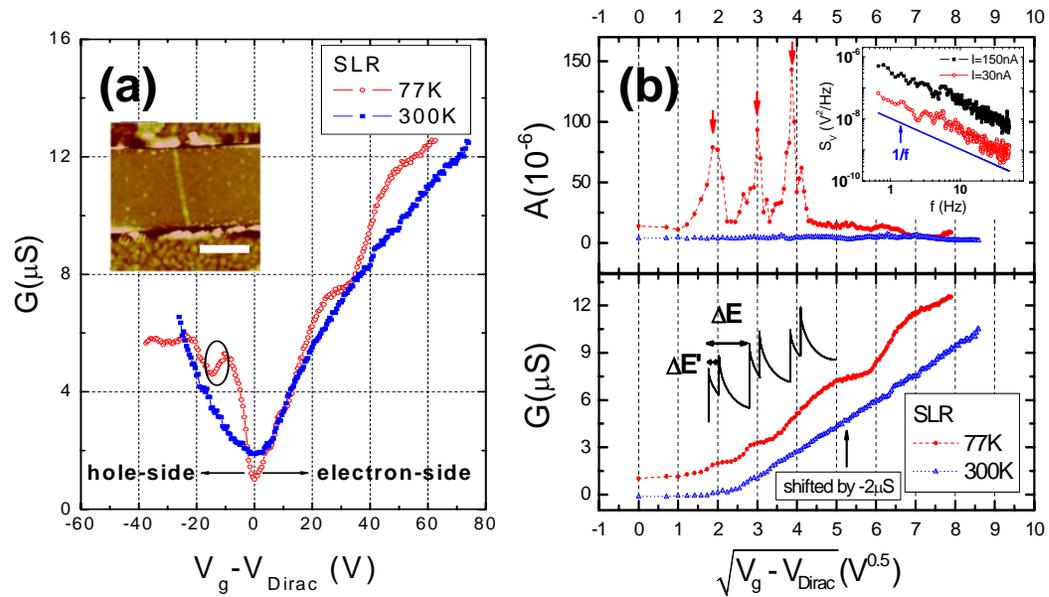

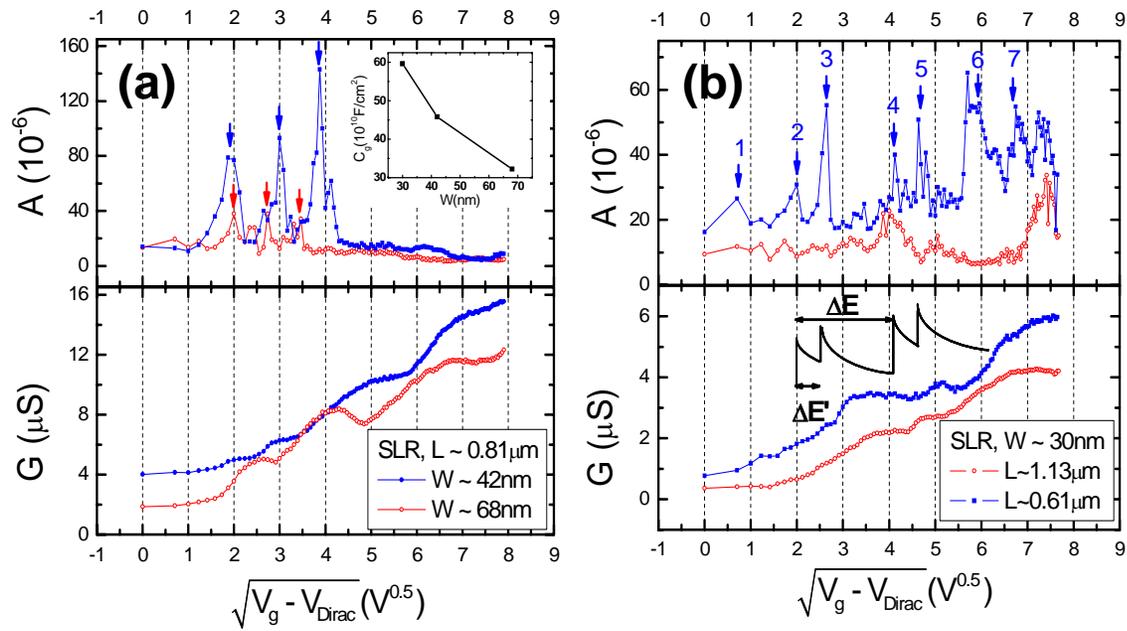

**Figure 3**
**Xu et. al.**

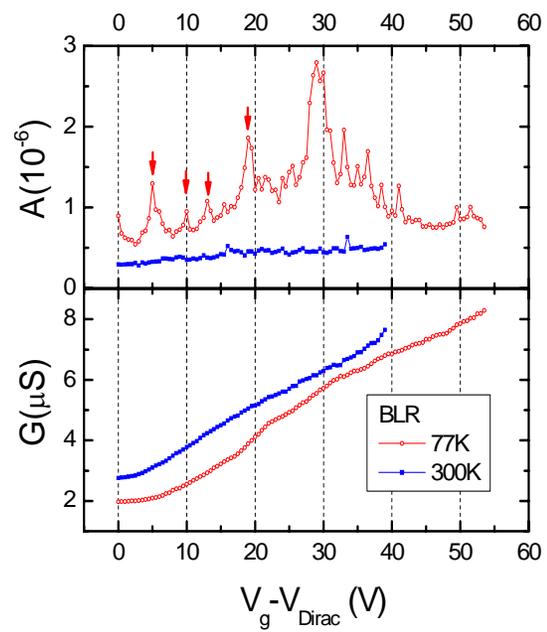

# Enhanced Conductance Fluctuation by Quantum Confinement Effect in Graphene Nanoribbons

*Guangyu Xu, Carlos M. Torres Jr. , Emil B. Song , Jianshi Tang, Jingwei Bai, Xiangfeng Duan ,*

*Yuegang Zhang, Kang L. Wang*

**Supplementary Materials**

I.  Methods

DEVICE PREPARATION

Bulk graphene sheets are mechanically exfoliated from natural graphite and transferred onto a 300 nm thermally grown $SiO_2$ dielectric film on highly doped Si substrates, and are identified through optical microscopy and Raman spectroscopy. Then, silicon nanowires are deposited onto the graphene sheets to serve as an etch mask (diameter ranging from 20nm-50nm) and located by means of an optical microscope (see the left part of Fig. S1)[S1]. A 30s oxygen plasma process is applied to etch away the exposed graphene, preserving the part of graphene (GNR) beneath the nanowires. The nanowires are removed by ultra-sonication, thus leaving only the GNR on the substrate (see the middle part of Fig. S1). Finally, the GNRs are patterned to multi-probe structures (see the right part of Fig. S1). An MMA/PMMA based dual-layer spin coating is applied followed by a 2 minute 150°C baking. After an e-beam lithography step, a Titanium/Gold metal layer is evaporated to serve as the electrical contacts with a thickness of 7nm/80nm, respectively. A 20 minute 100°C vacuum bakeout process is generally applied to partially desorb contaminants. The yield of this method exceeds 90% out of more than 70 fabricated GNR devices.

NOISE MEASUREMENT

a) The noise measurement is performed in a Janis 500 four-arm probe station, pumped down to a vacuum of $10^{-6}$ torr at room temperature. A copper sample stage is used as the back-gate and contacted by the cold finger. The graphene devices are fixed onto the stage through the double-sided copper tapes. The gate bias is added through the copper tape to the highly doped Si, where the series resistance is generally below 1kΩ. An Agilent 4156C is used to apply dc current bias to the device and to measure its dc conductivity. A SR560 is calibrated to amplify the voltage difference across the GNR sample, and an Agilent 35670A is used to measure the low frequency noise spectrum typically from f=0.05Hz to f=800Hz. The noise data are averaged 20 times from the fast Fourier transform (FFT) of the time-domain sampling data, and subtracted by the background noise, which was measured at zero current bias. The conductivity is averaged by 10 times of measurements via the 4156C at the same time of the noise measurement in order to avoid the inconsistency of the data caused by hysteresis effects. The temperature is controlled by a SI9700, and the vapor flow is adjusted by the needle valve of the transfer line. During the low temperature measurement, the temperature fluctuation is less than 4mK and the vacuum change is less than 3%.

b) We measured multiple GNR samples under different conditions to rule out extrinsic possibilities for the appearance of the noise peaks. These noise peaks were found to be independent of: 1) the types of the initial doping (i.e. the sign of $V_{Dirac}$); 2) the hysteresis effect caused by mobile ions close to the GNR-$SiO_2$ interface (the noise peaks are repeatable and independent of the gate sweeping directions); 3) the current biases ($A \sim S_V/V^2$ is not affected by the current-induced local heating effects)[S2, S3]; 4) the applied frequency ranges (1/f behavior is maintained for a broad frequency range). Overall, the noise peaks are universal in our SLR samples and suggest originating from the intrinsic trapping/de-trapping processes in their quasi-1D transport. The appearance of the noise peaks near the threshold of quasi-1D subbands has been observed in quantum point contact structures composed of III-V compound materials, where both the Lorentzian and 1/f noise spectra can exist depending on the samples[S4]. In our GNRs, we confirmed that the noise spectra follow 1/f behavior at each applied gate bias.

## II.    Definition of the noise figure

The noise in Ref. S5 is defined as the integration over the full frequency window, which reduces the measurement errors of the noise at specific frequencies[S1]. However, the original definition in Ref. S5 is frequency-dependent and may not rule out the noise spectra which deviate from the typical 1/f behavior. Hence, the noise in this work is given by $A = \frac{1}{N}\sum_{i=1}^{N} f_i \cdot S_{Vi}/V^2$, which takes the average over the frequency ranges where the noise follows with $1/f^\alpha$ behavior ($\alpha$ normally ranges from 0.85 to 1.12). This method might rule out other types of possibly pronounced noise (e.g. thermal noise, ac electricity power noise, etc).

## III.    AFM image of SLR1 and SLR2

We use detailed AFM scanning to compare the disorder of SLR1 (W~42nm) and SLR2 (W~68nm) in Fig. S2. The conductance of SLR2 (W~68nm) is lower than that of SLR1 (W~42nm) (even though W (SLR2) > W (SLR1)), possibly due to more disorder scattering in SLR2 (e.g. non-uniformity of the width and more MMA/PMMA residues). This fact shows that the disorder may affect the SLR transport significantly.

## IV.    K-K' subband splitting of SLR1 at T=4.19K

We performed measurements for SLR1 at T=4.19K (see Fig. S3), where more noise peaks (i.e. corresponding to the K-K' subband splitting, which originates from the chirality) are observed ($\Delta E' / \Delta E$ ~0.4) while the main peaks at 77K (see Fig. 2a in the main text) are maintained. That is, the noise peaks 1 and 3 appear at the same bias as T=77K, whereas the noise peak 2 is not observed at T=77K. This fact suggests that the K-K' subband splitting effect at 77K can be smeared by temperature. Device-dependent disorders may also smear the K-K' subband splitting; for example, noise peaks at higher-order subbands of SLR3 may not strictly follow the ratio $\Delta E'/\Delta E$ due to short-range disorder (see Fig. 2b in the main text).

V.     **Weak temperature-dependence of conductance of SLR and BLR**

Fig. S4 shows the T-dependence of dc conductance (G) for BLR1 at both electron- and hole-conduction sides. It is shown that BLR1 has no clear (or weak) T-dependence of G. The weak T-dependence of G is generally observed throughout our measurements for multiple SLR and BLR devices.

**References**

S1. Bai, J.; et al. Rational fabrication of graphene nanoribbons using a nanowire etch mask. *Nano Lett.* **2009**, 9, 2083-2087.

S2. Tan, Y.-W.; et al. Measurement of scattering rate and minimum conductivity in graphene. *Phys. Rev. Lett.* **2007**, 99, 246803.

S3. Barreiro, A.; et al. Transport properties of graphene in the high-current limit. *Phys. Rev. Lett.* **2009,** 103, 076601.

S4. Dekker, C.; et al. Spontaneous resistance switching and low-frequency noise in quantum point contacts. *Phys. Rev. Lett.* **1991**, 66, 2148-2151.

S5. Pal, A. N. and Ghosh, A. Resistance noise in electrically biased bilayer graphene. *Phys. Rev. Lett.* **2009**, 102, 126805.

**Supplementary Figures**

**Figure S1 Nanowire-mask based methods for GNR-FET fabrication. (Left)** Nanowire patterned onto single layer (light blue) and multilayer (dark blue) bulk graphene sheet on top of 300nm SiO$_2$ layer (purple). **(Middle)** AFM image of SLR (diameter~45nm, scale bar equals 0.3μm) after removal of the nanowire through ultra-sonication. **(Right)** Fabricated multi-probe SLR devices. Three nanoribbons were defined by the nanowire patterning. The distance between the two nearest Ti/Au electrodes is generally less than 500nm in short channel devices.

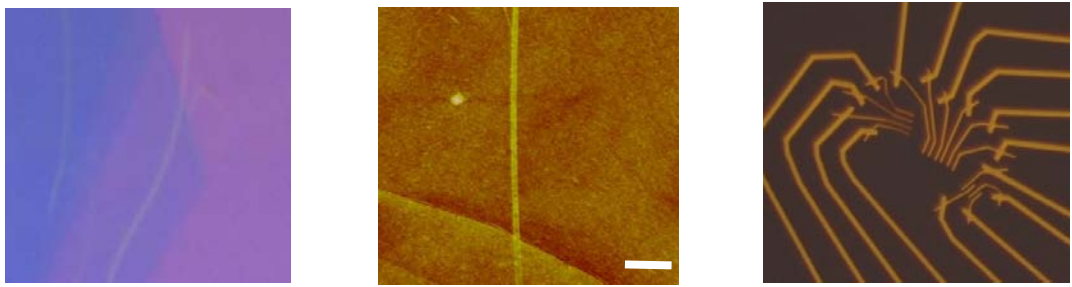

**Figure S2 Comparison of AFM image of SLR1 (left) and SLR2 (right). Scale bars equal to 0.3μm.**

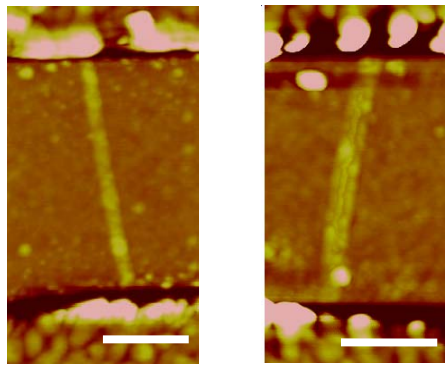

**Figure S3 Noise behavior of SLR1 at T=4.19K.** The noise peaks 1 and 3 appear at the same bias as T=77K, whereas the noise peak 2 is not observed at T=77K.

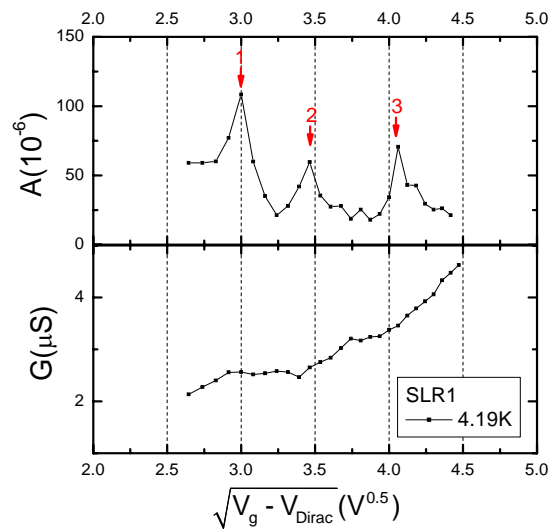

**Figure S4 Temperature-dependence of dc conductance for BLR1**

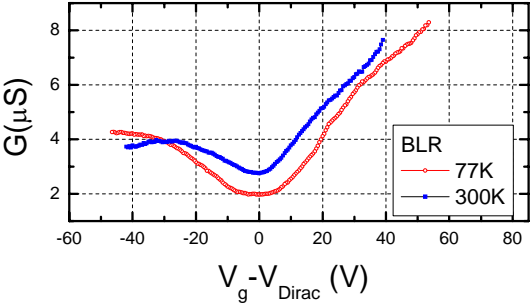